\documentclass[twocolumn,showpacs,a4paper,superscriptaddress,nofootinbib,tightenlines,floats]{revtex4} 
\usepackage{bm}
\usepackage{latexsym}
\usepackage{dcolumn}
\usepackage{amsfonts,amssymb}
\usepackage{graphicx}
\usepackage{psfrag}
\usepackage{amsmath}
\begin{document}

\def\apjl{Astrophys. J. Lett.}
\def\mnras{Mon. Not. Roy. Astron. Soc.}
\def\mnrasl{Mon. Not. Roy. Astron. Soc. Lett.}
\def\physrep{Phys. Rept.}
\def\apjs{Astrophys. J. Suppl.}
\def\aap{Astron. Astrophys.}
\def\araa{Annu. Rev. Astron. and Astrophys.}
\def\aj{Astron. J.}

\title{Evolution of perturbations in distinct classes of canonical scalar
  field models of dark energy}   

\author{H.~K.~Jassal}

\email[$^a$]{Email: hkj@hri.res.in}

\affiliation{Harish-Chandra Research Institute, Chhatnag Road,\\
Jhunsi, Allahabad 211 019, India.\\
}

\begin{abstract}
Dark energy must cluster in order to be consistent with the
equivalence principle.  
The background evolution can be effectively modeled by either a scalar field
or by a barotropic fluid.
The fluid model can be used to emulate perturbations in a scalar field model
of dark energy, though this model breaks down at large scales.
In this paper we study evolution of dark energy perturbations in
canonical scalar field models: the classes of thawing and freezing
models. 
The dark energy equation of state evolves differently in these classes.
In freezing models, the equation of state deviates from that of a
cosmological constant at early times.
For thawing models, the dark energy equation of state remains near that of the
cosmological constant at early times and begins to deviate from it only at
late times. 
Since the dark energy equation of state evolves differently in these classes,
the dark energy perturbations too evolve differently. 
In freezing models, since the equation of state deviates from that of a
cosmological constant at early times, there is a significant difference in
evolution of matter perturbations from those in the cosmological constant
model.  
In comparison, matter perturbations in thawing models differ from the
cosmological constant only at late times.
This difference provides an additional handle to distinguish between these
classes of models and this difference should manifest itself in the ISW
effect.  
\end{abstract}

\preprint{}

\maketitle

\section{Introduction}
Various observation have confirmed that the expansion rate of the universe
is accelerating \cite{obs_proof}. 
These observations include those of Supernova type Ia \cite{nova_data},
observations of Cosmic Microwave Background \cite{boomerang,wmap_params} and
large scale structure \cite{sdss}. 
The accelerated expansion of the universe can  be explained by
introducing a cosmological constant $\Lambda$ in the Einstein's equation
\cite{ccprob_wein,review3}.
However, the cosmological constant model is plagued by the fine tuning
problem \cite{ccprob_wein}.
This has motivated the study of dark energy models to explain the current
accelerated expansion of the universe (for reviews see \cite{DEreview}).
An alternative to the cosmological constant model is to
assume that this accelerated expansion is driven by a canonical scalar field
with a potential $V(\phi)$, namely the quintessence field
\cite{quint1,expo,linear,quadratic,invphi,invexpo}. 
There exists another class of string theory inspired scalar field dark energy
models known as tachyon models \cite{tachyon1,2003PhRvD..67f3504B} and there
are models which allow $w<-1$ are known as phantom models
\cite{2002PhLB..545...23C}. 
Phantom type dark energy can also be realized in a scalar tensor theory of
gravitation \cite{STG}. 
Other scalar field models include k-essence field \cite{2001PhRvD..63j3510A},
branes \cite{brane1} and fluid models like the Chaplygin gas model and its
generalizations \cite{chaply}. 
There are also some phenomenological models \cite{water}, field theoretical
and renormalization group based models  (see e.g. \cite{tp173}), models that
unify dark matter and dark energy \cite{unified_dedm1}, holographic dark
energy models \cite{HGDE}, QCD dark energy \cite{qcddark}  and many others like those based on horizon
thermodynamics (e.g. see \cite{2005astro.ph..5133S}).

Different models of dark energy which have the same background evolution are
indistinguishable purely from distance measurements.
Evolution of perturbations in these models is expected to break this
degeneracy.
The Integrated Sachs Wolfe (ISW) effect can distinguish a
cosmological constant from other models of dark energy, especially ones with a
dynamical dark energy \cite{ddw}. 
Dark energy perturbations have been extensively studied in the linear
approximation \cite{weller_lewis,bean_dore,depert,gordonhu}.
Perturbations in dark energy affect the low $l$ quadrupole in the CMB angular
power spectrum through the ISW effect \cite{weller_lewis,bean_dore}.
It was shown in Ref.\cite{weller_lewis} that dark energy perturbations affect
the low $l$ quadrupole in the CMB angular power spectrum through the ISW
effect.
For models with $w>-1$ this effect is enhanced while for phantom like models
it is suppressed.
In these models dark matter perturbations and dark energy perturbations are
anti-correlated for large effective sound speeds.
This anti-correlation is a gauge dependent effect
\cite{bean_dore}.
There are several other studies of perturbations in dark energy
\cite{chpgas_pert}, including some that deal with evolution of
spherical perturbations \cite{sph_coll,mota}. 
For canonical scalar field dark energy, the perturbations in matter are
enhanced by the presence of dark energy perturbations in comparison with
smooth dark energy model \cite{ujs}. 
The matter perturbations in fluid models are suppressed compared to
corresponding homogeneous dark energy scenario \cite{hkj}.
As long as the speed of propagation of perturbations `$c_s^2$' is positive
the evolution of matter perturbations is indistinguishable from a smooth dark
energy model.  
This is true for scales smaller then the Hubble radius.
Dark energy perturbations in a fluid model with an appropriate $c_s^2$ emulate
that of a scalar field model very well below the Hubble scale but start to
differ at larger scales.
Therefore the fluid model is not a good approximation at these scales.
This also implies that the growth of perturbations at large scales depends on
the details of the model even though the background evolution is the
same.
A separate analysis is therefore required for every  model.

In this paper we consider different scalar field models to study evolution of
dark energy perturbations. 
We consider two different types of potentials, classified as '{\it thawing}' and
'{\it freezing}' in Ref. \cite{limitsofq}.
For potentials with a thawing behavior, the scalar field is frozen at early
times and starts to roll down the potential at late time.
Hence the equation of state of dark energy starts near $w=-1$ at early
times to $w>-1$ at late times. 
In contrast, in the case of potentials with freezing behavior, the scalar field
rolls down the potential and approaches the minimum of the potential, with the
equation of state going from $w>-1$ to freezing at $w=-1$ at late times.
Due to the different evolution of the equation of state, it is expected that the
perturbations in dark energy evolve differently in these classes. 
For freezing potentials, at early times, the equation of state deviates from
$w=-1$ hence the perturbations in dark energy are expected to be enhanced. 
Whereas, in thawing potential scenarios, dark energy perturbations become
significant only at late times.
Since dark energy perturbations enhance perturbations in nonrelativistic
matter, models with  freezing type potentials will have more enhanced
perturbations in matter  than those with thawing type behavior.
This can be an additional tool (apart from distance
measurements) to distinguish between these types of models.

The paper is organized as follows. In Section \ref{sec::fg} we discuss
evolution of perturbations in scalar field potentials in the thawing and
freezing classes.
The results are summarized in the concluding Section \ref{sec::concl}.

\section{Perturbations in scalar field dark energy} \label{sec::fg}

To describe dark energy perturbations, we choose the  Newtonian gauge.
We choose to work in this gauge as in this case we can directly relate the
metric perturbation $\Phi$ to the gravitational potential perturbation. 
In the absence of anisotropic stress, the perturbed metric can be
written in the form   
\begin{equation}
ds^2 = (1+ 2\Phi) dt^2 -  a^{2}(t) \left[(1- 2\Phi) \delta_{\alpha \beta}
  dx^{\alpha} dx^{\beta}\right] 
\end{equation}
where $\Phi$ is the gauge invariant potential defined in \cite{bardeen}. 
\

The linearized Einstein equations obtained from the above metric are given by
\begin{eqnarray} \label{eq::einstein}
\frac{k^2}{a^2} \Phi + 3 \frac{\dot{a}}{a} \left(\dot{\Phi} +
  \frac{\dot{a}}{a} \Phi \right) &=& \\ \nonumber
&-& 4\pi G  \left[\rho_{NR} \delta_{NR} +
  \rho_{DE} \delta_{DE} \right] \\ \nonumber 
\dot{\Phi} + \frac{\dot{a}}{a} \Phi &=& -4 \pi G \left[\rho_{NR} v_{NR} +
  \rho_{DE} v_{DE}\right] \\
\nonumber 
4 \frac{\dot{a}}{a} \dot{\Phi} + 2  \frac{\ddot{a}}{a} \Phi +
\frac{\dot{a}^2}{a^2} \Phi + \ddot{\Phi} &=& 4 \pi G  \delta P
\end{eqnarray}
where dot denotes derivative with respect to the coordinate time $t$ 
and $v_{NR}$ is the potential for the matter peculiar velocity such that
$\delta u_{i} = \nabla_{i}v_{NR}$. 
In these equations, we have Fourier decomposed the perturbed
quantities such as $\Phi$, $\delta \phi$,  $\delta \rho_{NR}$ and $v_{NR}$ and
replaced $\nabla^{2}$ by $-k^{2}$, where $k$ is the wave number defined as $k
= 2\pi/\lambda$  with $\lambda$ being the comoving length scale of
perturbation.
In these equations all the perturbed quantities  correspond to the amplitude
of perturbations in the $k^{th}$ mode.

\begin{figure*}
\begin{center}
\begin{tabular}{cc}
\includegraphics[width=0.5\textwidth]{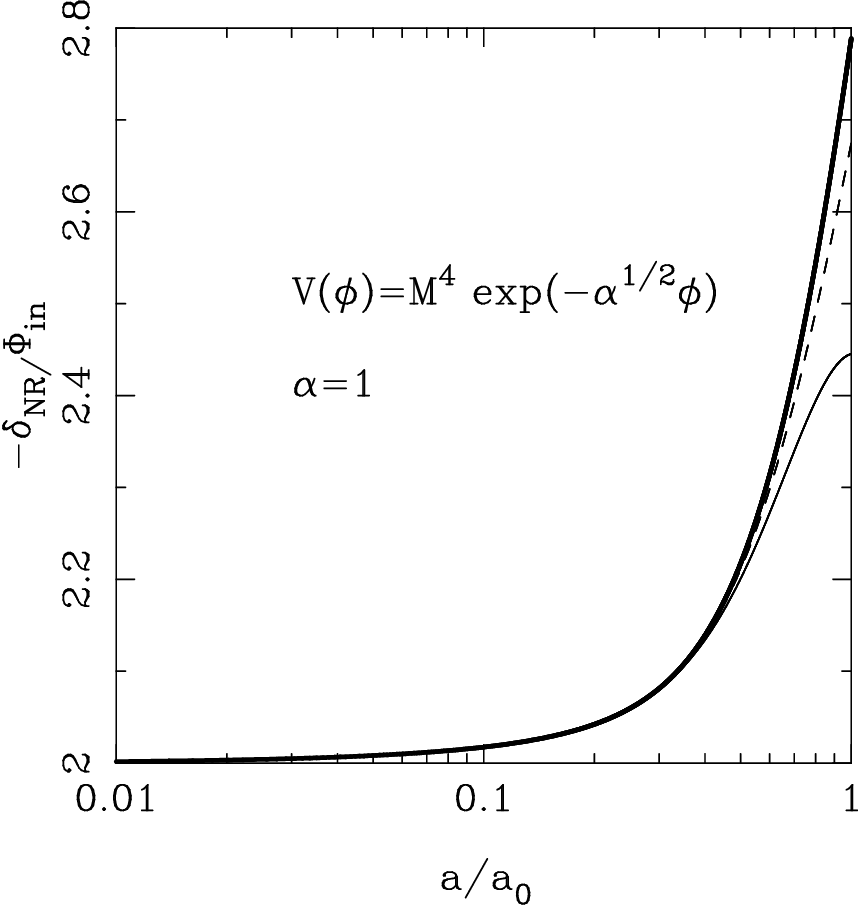} 
\includegraphics[width=0.5\textwidth]{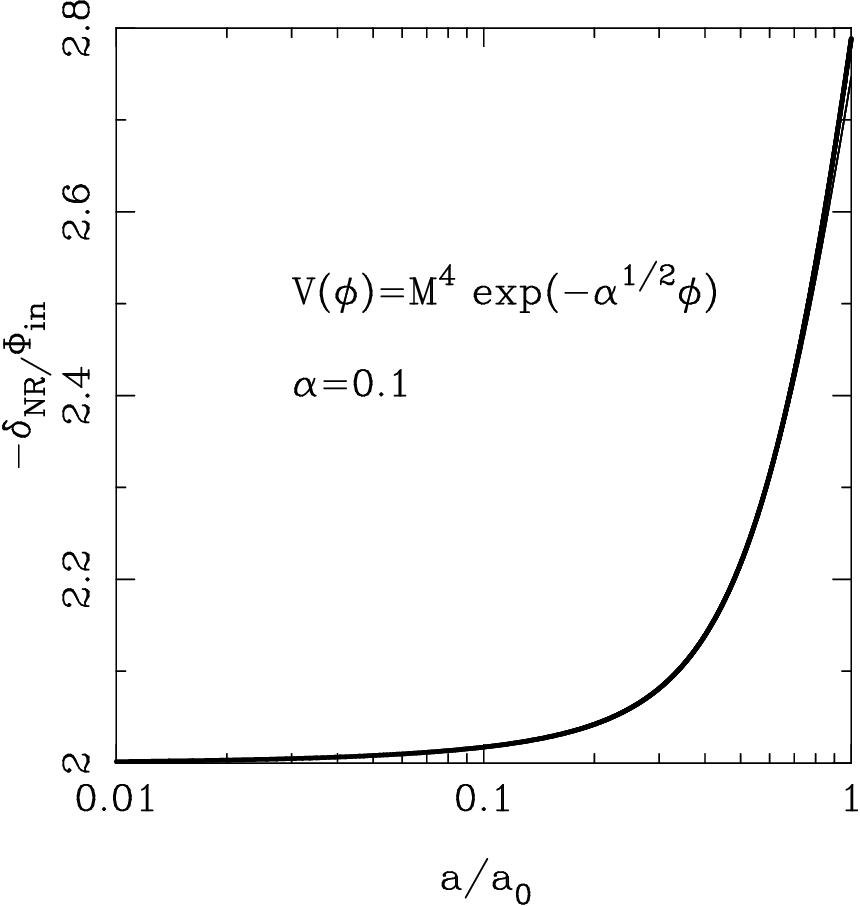}
\end{tabular}
\end{center}
\caption{The figure shows evolution of nonrelativistic matter density contrast
as a function of the scale factor for the exponential potential (TH1).  
The thick solid line is the evolution of density contrast in the cosmological
constant model.
The solid line is for a homogeneous dark energy scenario whereas the dashed
line show the evolution if dark energy perturbations are taken into
account.}      
\label{fig::expo}
\end{figure*}

We assume a flat spatial cosmology and we choose the initial equation of state
$w$ to be very close to $-1$ at early times (say redshift of $z_{i} = 100$).  
The initial value of the scalar field needs to be fine tuned such that
universe begins to accelerate at late times.
We fine tune the parameters such that the matter density parameter 
is within the range allowed by the present day observations.
For a system of pressureless matter and a scalar field, the dynamics
of perturbations is uniquely determined by $\Phi(t)$  and $\delta \phi(t)$. 
Hence we need two second order equations which connect $\Phi(t)$ and $\delta
\phi(t)$.
As one of these equations, we choose the third equation in system
\ref{eq::einstein}. 
The dynamical equation for the perturbations in the scalar field $\delta
\phi(t)$ is given by :  
\begin{equation}
\ddot{\delta \phi} + 3\frac{\dot{a}}{a}\dot{\delta \phi} + \frac{k^2
  \delta \phi}{a^2} + 2\Phi V'(\phi)  -4\dot{\Phi}\dot{\phi} +
V''(\phi)\delta \phi = 0 \label{eqn::pertkg}
\end{equation}

These equations complete the system and we can then calculate the fractional
density perturbation for dark energy and for nonrelativistic matter as
\begin{eqnarray}
\delta_{\phi} &=& \frac{1}{ \frac{1}{2}\dot{\phi}^2 +
  V(\phi)}\Big[\dot{\phi} \dot{\delta\phi} - \Phi\dot{\phi}^2 +
  V'(\phi)\delta{\phi}\Big]\label{eqn::deltaphi}\\
\delta_{NR} &=& -\frac{1}{4 \pi G \rho_{NR_{o}}a^{-3}}\Big
  \{3\frac{\dot{a^2}}{a^2} \Phi +3 \frac{\dot{a}}{a}\dot{\Phi} +
  \frac{k^2\Phi}{a^2} \Big \} \\ \nonumber 
&+&
  \frac{\delta_{\phi}}{\rho_{NR_{o}}a^{-3}}\Big[ \frac{1}{2}\dot{\phi}^2 
  + V(\phi)\Big]\label{eqn::deltaNR}
\end{eqnarray}

To solve the equations numerically, we introduce the following 
dimensionless variables  
\begin{equation}
\Phi_{N} = \frac{\Phi}{\Phi_{\mathrm{in}}},~~~~~\delta y = \frac{\delta
  \phi}{\Phi_{\mathrm{in}}M_{p}} 
\label{eqn::redef_del_phi}
\end{equation}
Here $\Phi_{N}$ is the normalized gravitational potential with
$\Phi_{\mathrm{in}}$ 
being the value of the metric potential at the initial time $t = t_{\mathrm{in}}$. 
In terms of these two new variables, the equations are
\begin{eqnarray}
\Phi_{_{N}}'' &+& 4\frac{s'}{s}\Phi_{N}' + \Big (2\frac{s''}{s} +
\frac{s'^{2}}{s^{2}}\Big )\Phi_{N} \\ \nonumber
&-& \frac{1}{2}\Big [y'\delta y'
-\Phi_{N}y'^{2} +  V''(y) \delta y \Big] = 0 \label{eqn::Finaleqn3} 
\end{eqnarray}
\begin{eqnarray} 
\delta y'' &+&  3\frac{s'}{s}\delta y' - 4\Phi_{N}'y' + \frac{\bar{k}^{2}\delta
  y}{s^{2}} \\ \nonumber 
&+& 2 \Phi_{N} V'(y) + V''(y) y' =0
\label{eqn::Finaleqn4}
\end{eqnarray}
Here $\bar{k} = kc/H_0$, is the wave number scaled with respect to the
Hubble radius,  $H_0$ being the Hubble parameter at the present epoch. 
The prime denotes derivative with respect to the variable $t H_0$.

\begin{figure*}
\begin{center}
\begin{tabular}{cc}
\includegraphics[width=0.5\textwidth]{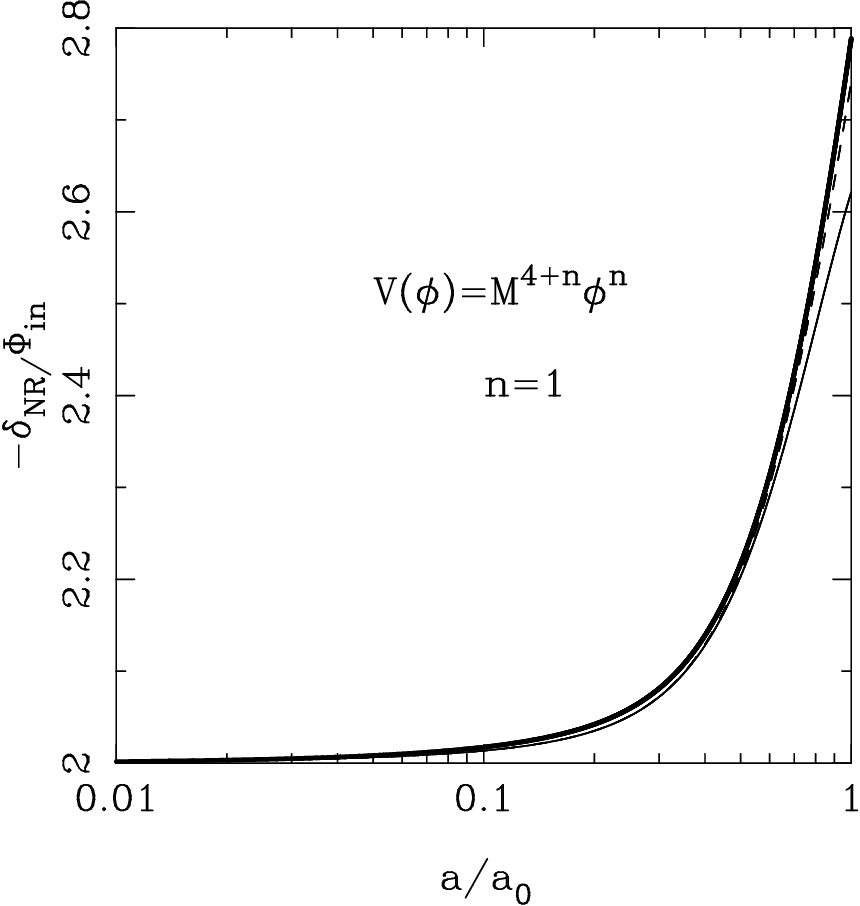} 
\includegraphics[width=0.5\textwidth]{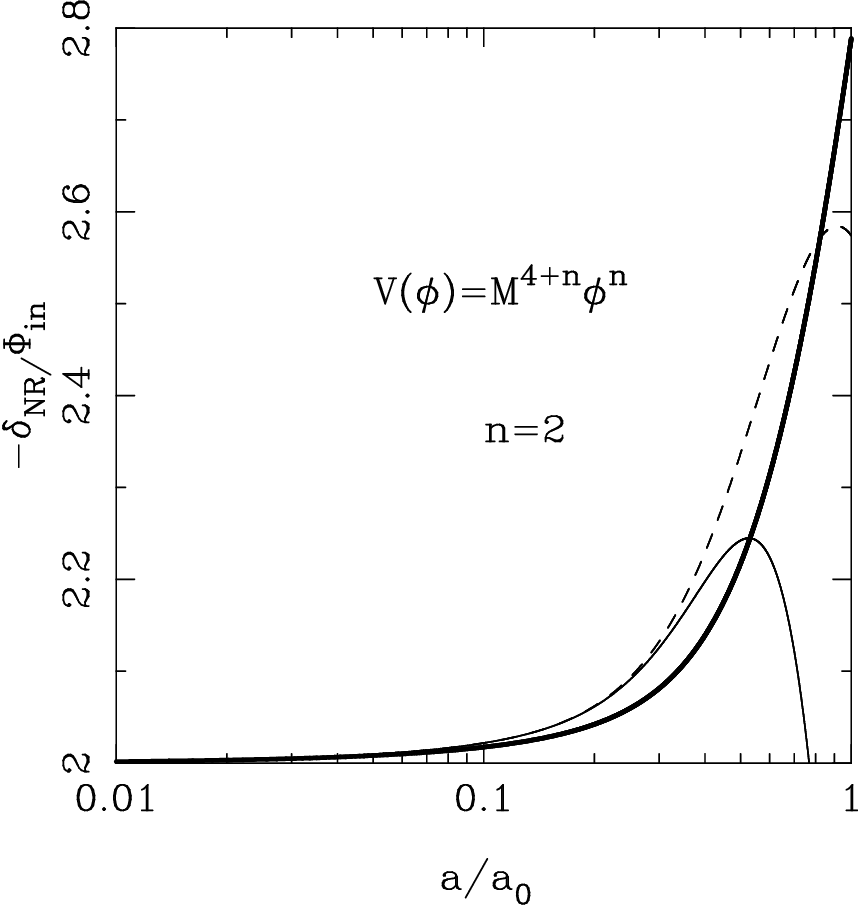}
\end{tabular}
\end{center}
\caption{The plot on the left shows evolution of matter density contrast for
  the concave potentials (TH2) with respect to the scale factor. 
  The figure on the left shows evolution of
  nonrelativistic matter density contrast as a function of the 
  scale factor for the linear potential. The figure on the right shows
  evolution of nonrelativistic matter density contrast as a function of the
  scale factor for the quadratic potential. The line styles are the same as in
Fig. 1.} 
\label{fig::phin}
\end{figure*}

We assume that the perturbation in the scalar field in the matter
dominated epoch is negligibly small compared to other
perturbed quantities such as $\Phi$ and $\delta_{NR}$.
Hence we can treat the scalar field to be initially homogeneous.
This corresponds to setting the initial condition $\delta y_{\mathrm{in}} = 0$
and $\delta y'_{\mathrm{in}} = 0$.
The only initial condition that needs to be determined is the value of
$\Phi'_{N_{\mathrm{in}}}$ at $ t = t_{\mathrm{in}}$.
In the matter dominated epoch, the gravitational potential
$\dot{\Phi}(t)=0$ for all values of the wave number $k$, therefore we can set
the initial condition $\Phi'_{N_{\mathrm{in}}}(k) = 0$.

If dark energy is a cosmological constant, then it does not cluster and the
gravitational field $\Phi$ decays when the cosmological constant dominates. 
For $w \neq -1$ too, the potential $\Phi$ remains constant in the matter
dominated era and starts to decay when dark energy contribution becomes
significant. 
This rate of decay is dependent on the value of $w$.
For a canonical scalar field, dark energy perturbations are correlated with
the matter perturbations and they enhance matter perturbations \cite{ujs}. 
If dark energy is a barotropic fluid, the matter perturbations are suppressed
as compared to the corresponding homogeneous model \cite{hkj}.
This difference is due to the fact the homogeneous limit is achieved
differently for the two models. 
For a scalar field,  even if one ignores spatial gradients there still remains
a contribution to $\delta P$. 
This contribution cancels with a corresponding contribution from the pressure
term due to the background. 
This term cancels for homogeneous scalar field dark energy and leads to a
suppression in matter perturbations.
For fluid models $\delta P$  vanishes and the residual pressure term due to
the background evolution makes the evolution different from that in a scalar
field. 
Therefore, assuming dark energy to be homogeneous leads to a large difference
in matter density contrast.

For sub-Hubble scales, a fluid model effectively emulates a scalar field
model \cite{hkj}.  
At scales smaller then the Hubble radius, dark energy can be assumed to be
homogeneous.
It is sufficient to assume a fluid approximation and the number counts of
clusters at different redshifts \cite{nm} distinguish deviation from
the cosmological constant \cite{sph_coll}. 
At larger scales, where dark energy perturbations may play a significant role,
the fluid analogy breaks down and the evolution of matter density contrast
depends on individual scalar field models.
This also implies that the growth of perturbations at large scales depends on
the details of the model even though the background evolution is the
same. 
A separate analysis is therefore required for every  model.
Since dark energy perturbations are significant at large scales, i.e., scales
larger then Hubble radius, we will restrict the subsequent discussion to large
scales.
In particular, the reference scale we choose here is the length scale
$\lambda=10^5$ Mpc.

The scalar field models have been classified as freezing or thawing in
\cite{limitsofq}.  
The classification is based on the difference in the way equation of state of
dark energy evolves.
The freezing type models are fast roll models with a steep potential.
The field remains subdominant and at late times becomes dominant and drives
the acceleration of the expansion of the universe.
The thawing models have a potential which is nearly flat.
Therefore, the equation of state starts with the value $w=-1$ at early times
and the energy density of the field is negligible.
As the universe expands, the energy density of the field becomes comparable
to the background energy density.
The equation of state of dark energy deviates away from its frozen value,
hence the scalar field thaws.

The dark energy equation of state evolves differently in these classes of
models. 
Therefore, the evolution of perturbations in freezing type and thawing type
models is expected to be different.
Using present day distance measurement observations, it is not possible to
distinguish between different dark energy models from The $\Lambda$CDM model
and also between various scalar field models.
These observations include Present day Supernova observations and data from
the Baryon Acoustic Oscillations measurements \cite{anjan}.
This degeneracy is expected to be removed  if dark energy perturbations are
included.
The difference in these models is expected to make a significant contribution
to integrated effects such as the ISW effect \cite{ddw}.

\begin{figure*}
\begin{center}
\begin{tabular}{cc}
\includegraphics[width=0.5\textwidth]{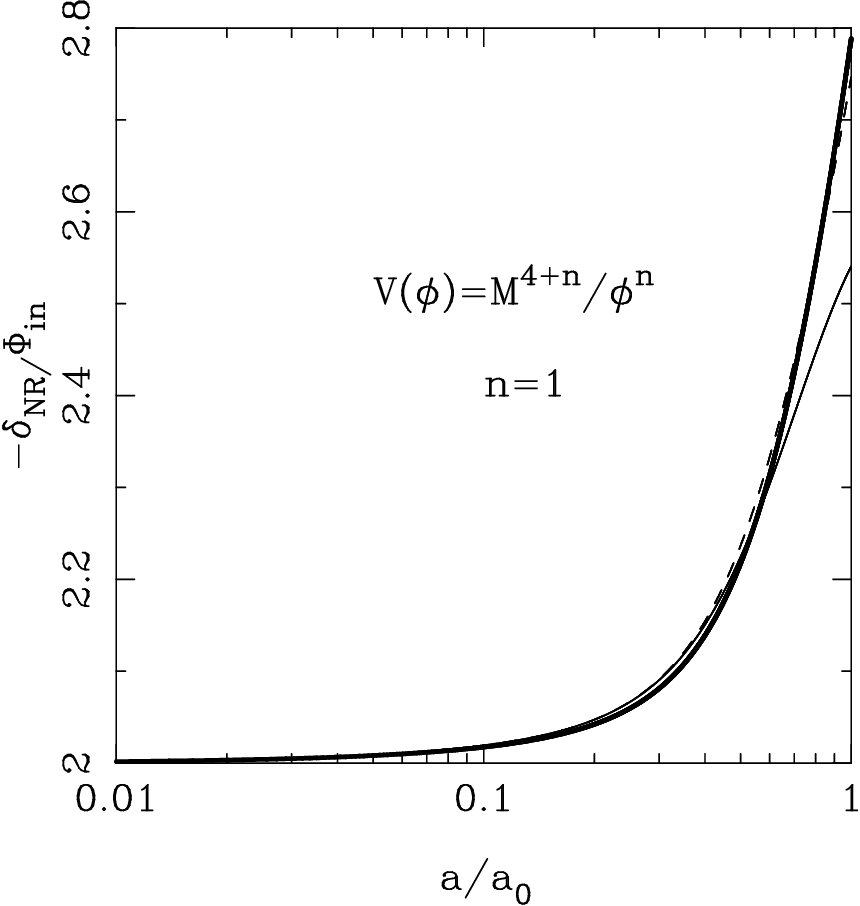} 
\includegraphics[width=0.5\textwidth]{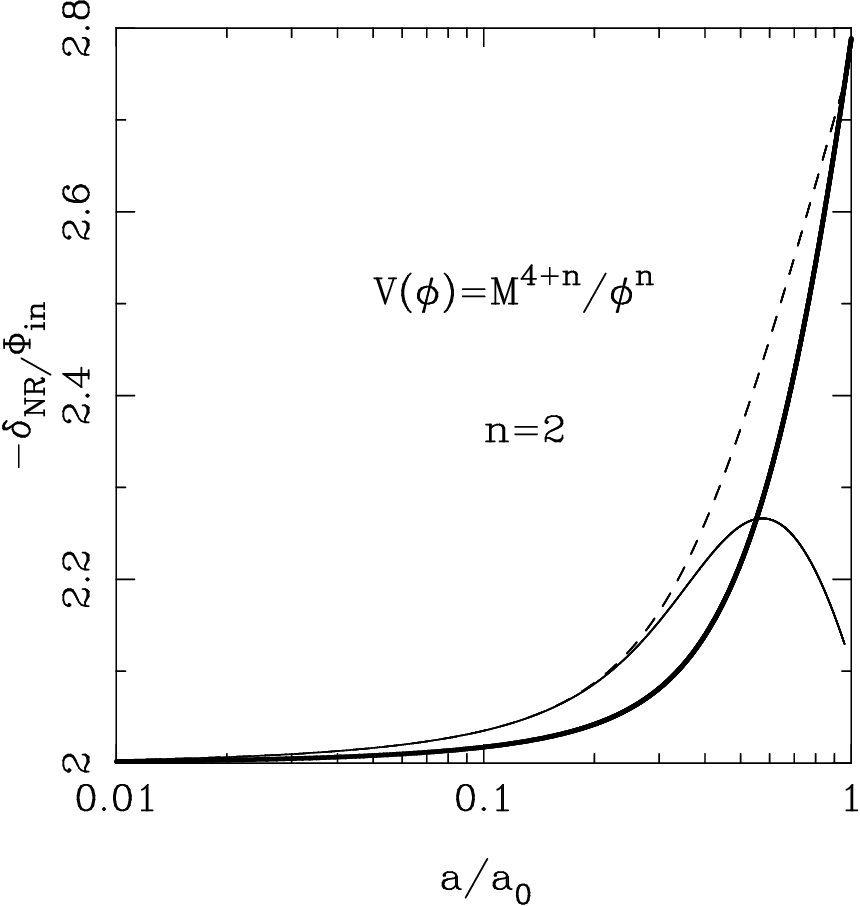}
\end{tabular}
\end{center}
\caption{The figure shows evolution of nonrelativistic matter density contrast
as a function of the scale factor for the inverse $\phi$ potentials (FR1) for
the parameter values mentioned.}
\label{fig::invphi}
\end{figure*}

To study scalar field perturbations, we consider the following potentials: 
\begin{enumerate}
\item For 'thawing' behavior
\begin{enumerate}
\item TH1: Exponential potential : \\ 
$V(\phi) = M^4 exp(-\sqrt{\alpha} \phi/M_P)$ \cite{expo} 
\item TH2: Polynomial (concave) potential :  \\ $V(\phi) = M^{4-n} \phi^n$
  \cite{linear,quadratic}
\end{enumerate}
\item For 'freezing' behavior
\begin{enumerate}
\item  FR1: Inverse power potential : \\ $V(\phi) = M^{4+n} \phi^{-n}$ \cite{invphi}
\item  FR2: $V(\phi) = M^{4+n} \phi^{-n} exp(\alpha \phi^2/M_P^2)$ \cite{invexpo}
\end{enumerate}
\end{enumerate}
We have labeled the scalar field potentials as TH (thawing) and FR (freezing),
and henceforth will be using these labels to refer to the models.
Since Einstein equations imply  homogeneous dark energy to be inconsistent, we
will compare the evolution of perturbations in  scalar field dark energy
with the cosmological constant. 
We have however also considered the homogeneous limit in all these models.

The growth of perturbations in scalar field model with an exponential
potential (TH1) has been studied in \cite{ujs}. 
For this potential, if $\alpha=1$, the present day equation of state is
$w=-0.86$ and the acceleration of the universe starts at redshift $z=0.81$.    
For smaller $\alpha$ the models $w$ is smaller, for instance if $\alpha =
0.1$, the equation of state at present is $w=-0.95$. 
In Fig. \ref{fig::expo}  we plot the evolution of matter perturbations at
scale $\lambda=10^5$ Mpc as a function of the scale factor $a$.
For smaller $\alpha$ where the present equation of state is closer to that of
a cosmological constant, there is no significant difference if  we assume dark
energy to be homogeneous or if we assume dark energy to be clustered.
If $\alpha=1$ there is a significant enhancement of matter perturbations
\cite{ujs}.
Compared to the cosmological constant model, matter perturbations in this
model are suppressed, with approximately $6$\% deviation.

In Fig. \ref{fig::phin} we show the evolution of matter density contrast for
the concave potential (TH2) with $n=1$ and $n=2$. 
The present day of equation of state is $w \approx -0.94$ if we choose  $n=1$
and reaches  $w \approx -0.3$  for $n=2$.
If the potential is quadratic, the gravitational potential initially decays,
and in the future starts to oscillates.
This is due to the fact that the scalar field reaches the minimum of the
potential and the equation of state for dark energy then oscillates between
$w=-1$ and $w=1$.
These oscillations are evident at earlier times at  very large scales, say at
$\lambda=10^5$ Mpc. 
At large scales, there is a significant enhancement in matter perturbation as
compared to the case when dark energy is assumed to be homogeneous.
In contrast, the gravitational potential $\Phi$ continues to decay if $n=1$.  
For $n=1$ the matter perturbations are suppressed as compared to the potential
with $n=2$ but they continue to grow and become larger when oscillations set in
the case of the quadratic potential. 
Matter perturbations in linear potential follow those in cosmological constant
model very closely.
In quadratic scalar field potential case, these perturbations are enhanced
compared to the $\Lambda$CDM model at early times and are suppressed at late
times.
At $a\approx 0.4$, the percentage enhancement as compared to the cosmological
constant model is $\approx 5$ \%.

\begin{figure*}
\begin{center}
\begin{tabular}{cc}
\includegraphics[width=0.5\textwidth]{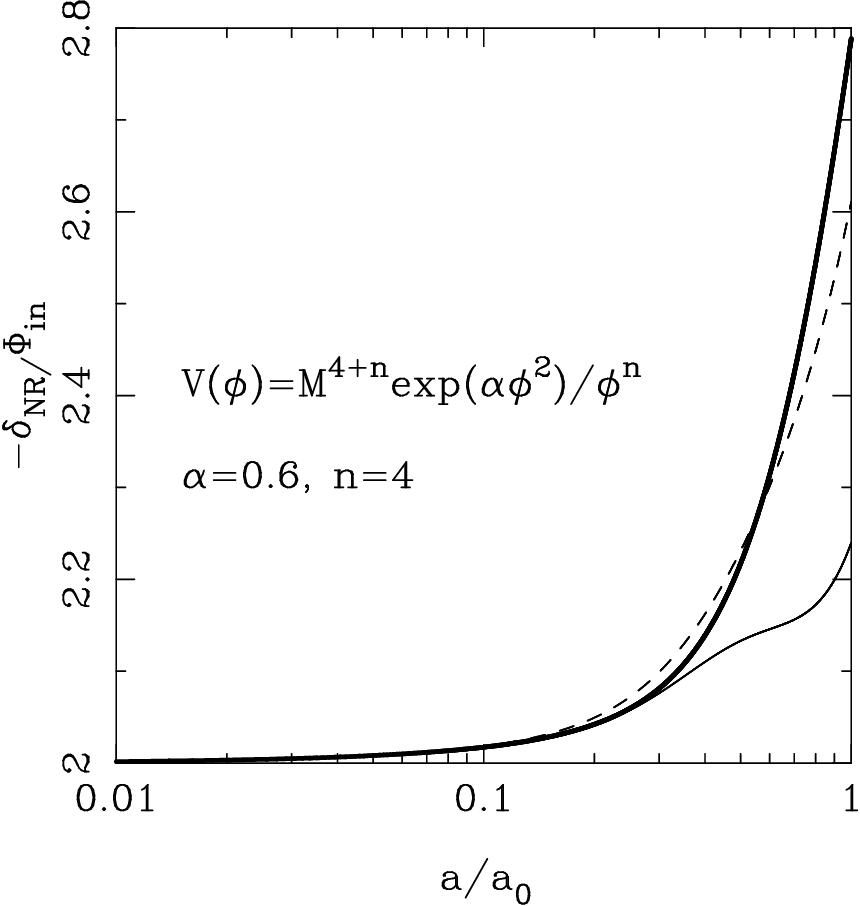} 
\includegraphics[width=0.5\textwidth]{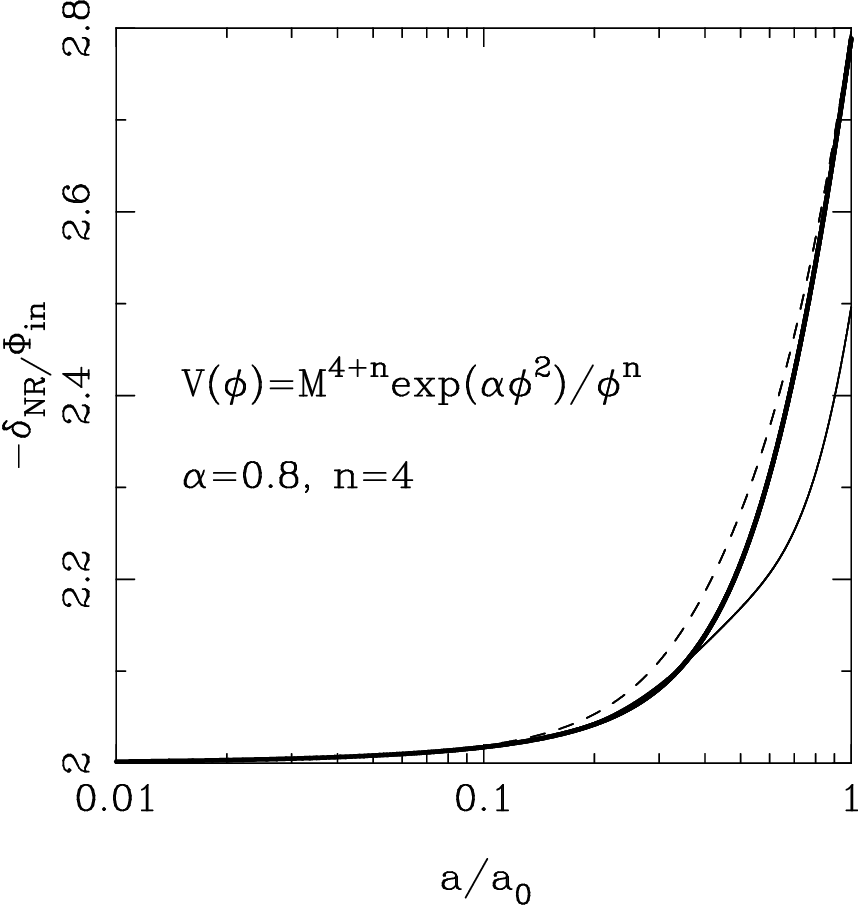}
\end{tabular}
\end{center}
\caption{The figure shows evolution of nonrelativistic matter density contrast
as a function of the scale factor for the potential $V(\phi) = M^{4+n}
\phi^{-n} exp(\alpha \phi^2/M_P^2)$  (FR2). The plot of the left corresponds to
$\alpha=0.6$ and $n=4$ . The plot  on the right shows the density contrast
for $\alpha=0.8$ and $n=4$.}
\label{fig::invphiexp}
\end{figure*}

For potential FR1, the scalar field begins to roll down the potential at early
times and begins to freeze at late times. 
In this freezing type evolution, the equation of motion of dark energy starts
away from $w=-1$ and approaches this value as the scalar field slows down. 
The epoch at which this freezing occurs is subject to fine tuning of
parameters.
If the scalar field is frozen before the present time, the duration of the
matter dominated phase is then too small, i.e., scalar field approaches the
kinetic energy dominated phase before reaching $z=1000$. 
To achieve a sufficiently prolonged matter dominated phase, the scalar field
should freeze in far future.
The dark energy equation of state reaches a maximum of approximately $-0.7$,
and we tune the initial value of $\phi$ such that we achieve $\Omega_{NR}
\approx 0.3$ at present. 
In Fig. \ref{fig::invphi} we show the evolution of matter density contrast with
scale factor.
There is an enhancement in matter perturbations with an increasing value of
$n$.
We have plotted the density contrast for  $n=1$ and for $n=2$. 
In this model, since the value of equation of state deviates away from $w=-1$
at early times, there is a significant enhancement in dark energy
perturbations and hence in matter perturbations.
With $n=1$, perturbations in nonrelativistic matter are comparable to those in
cosmological constant model, and with $n=2$ they are enhanced by approximately
$7$ \% at $a \approx 0.3$ with respect to the cosmological constant model.

For the potential FR2, the freezing behavior is achieved earlier without
trading away a viable duration of the matter dominated era.
For a given $n$, the present value of the equation of state moves further
away from $w=-1$ as $\alpha$ increases.
Also for a constant $\alpha$, the value of the equation of state increases
with an increase in $n$.
As in all other models, at very early time we assume $w=-1$.
For instance, if $\alpha=0.6$, for $n=1$ reaches $ w \approx -0.9975$ at
redshift $\approx 0.4$ and then starts to approach the freezing behavior.
If $\alpha=0.8$ and $n=4$, the maximum  value the equation of state reaches is
$w=-0.82$.
In Fig. \ref{fig::invphiexp}, we show the evolution of matter density
contrast at $\lambda=10^5$ for two different parameter sets and also in
comparison to that in $\Lambda$CDM model.

\begin{figure*}
\begin{center}
\begin{tabular}{cc}
\includegraphics[width=0.5\textwidth]{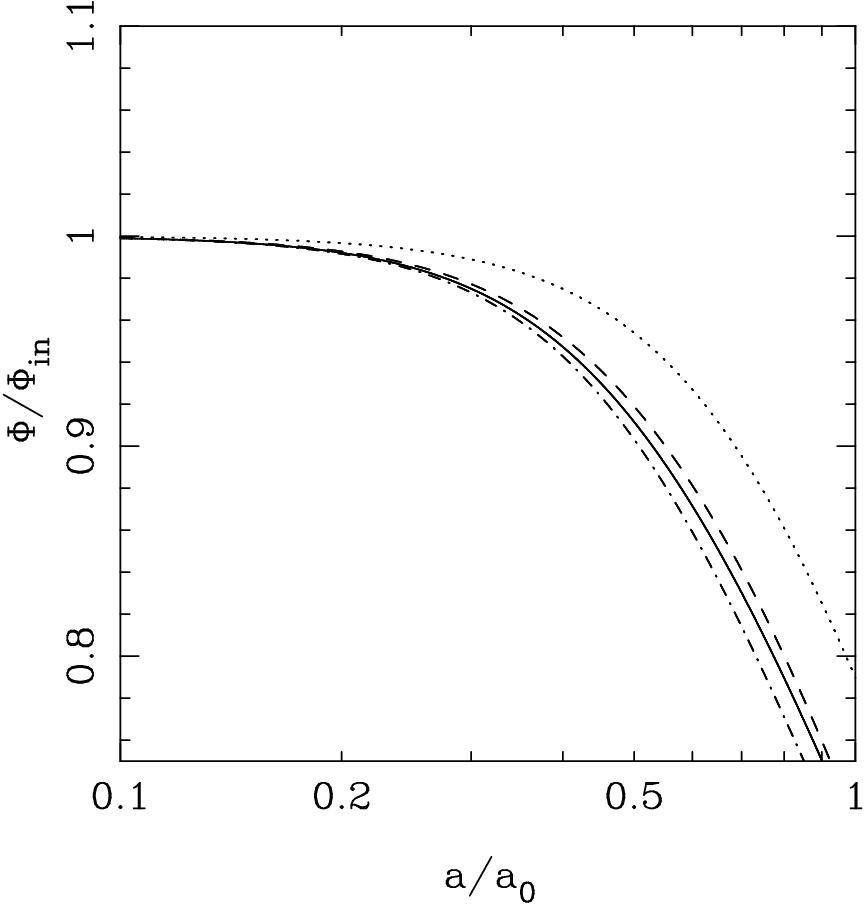}
\includegraphics[width=0.5\textwidth]{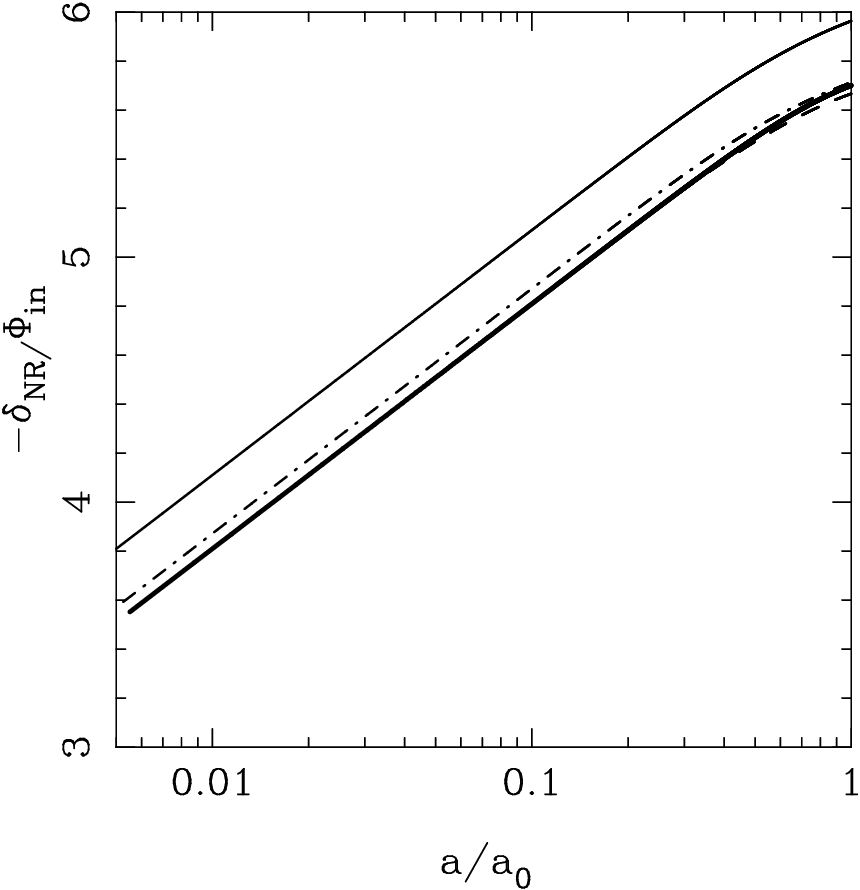}
\end{tabular}
\end{center}
\caption{This figure shows evolution of the gravitational
  potential $\Phi$   and  matter density contrast $\delta_{NR}$ as a
  function of scale factor for the  four different models at length scale
  $\lambda=50$ Mpc. The solid line shows the evolution for the scalar field
  potential $V(\phi)=M^{4+n} \phi^{-n}$ (FR1) with $n=2$. 
The dashed line is for the
  potential $V(\phi)=M^{4+n} \phi^{-n} exp(\alpha \phi^2/M_P^2)$ (FR2) with
  parameters $\alpha=0.8$ and $n=4$. 
The
dotted line corresponds to the thawing exponential potential $V(\phi) = M^4
exp(-\sqrt{\alpha} \phi/M_P)$ (TH1) where $\alpha=1$ and the dot dashed line
corresponds to the polynomial potential $V(\phi)=M^{4-n} \phi^{n}$ (TH2) for
$n=2$. At 
early times, the 
freezing potentials have matter perturbations enhanced as compared to the
thawing potentials. The evolution at this scale is same for a homogeneous dark
energy and for a clustering dark energy. The thick solid line corresponds to
the $\Lambda$CDM model.}  
\label{fig::delta1}
\end{figure*}

\begin{figure*}
\begin{center}
\begin{tabular}{cc}
\includegraphics[width=0.5\textwidth]{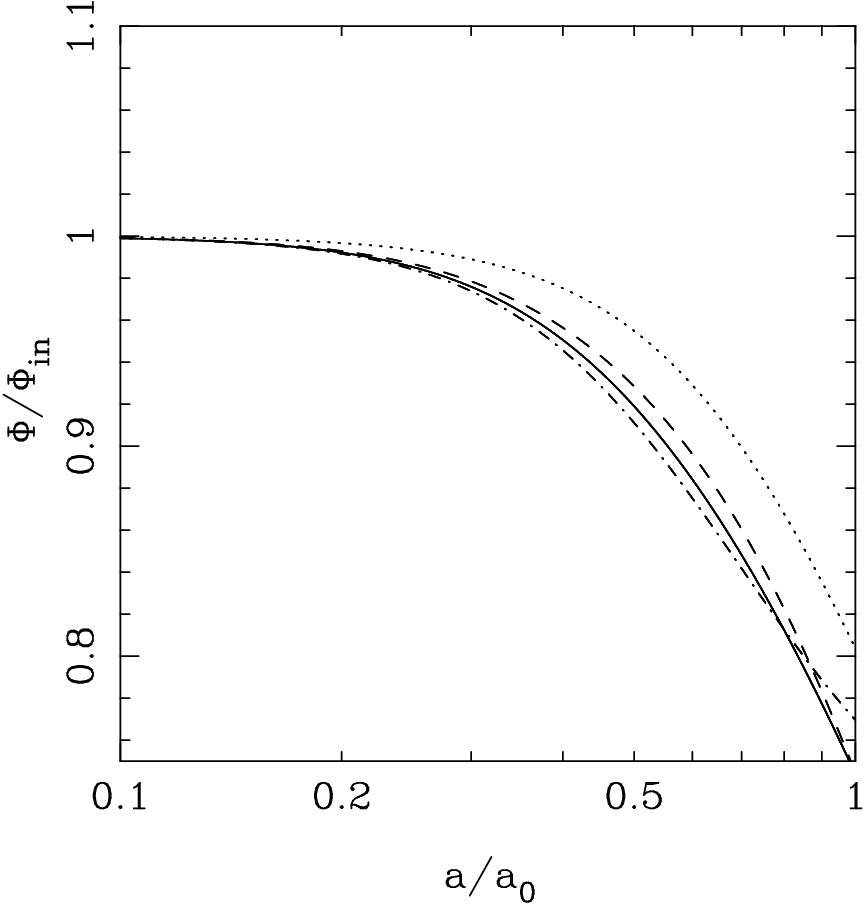}
\includegraphics[width=0.5\textwidth]{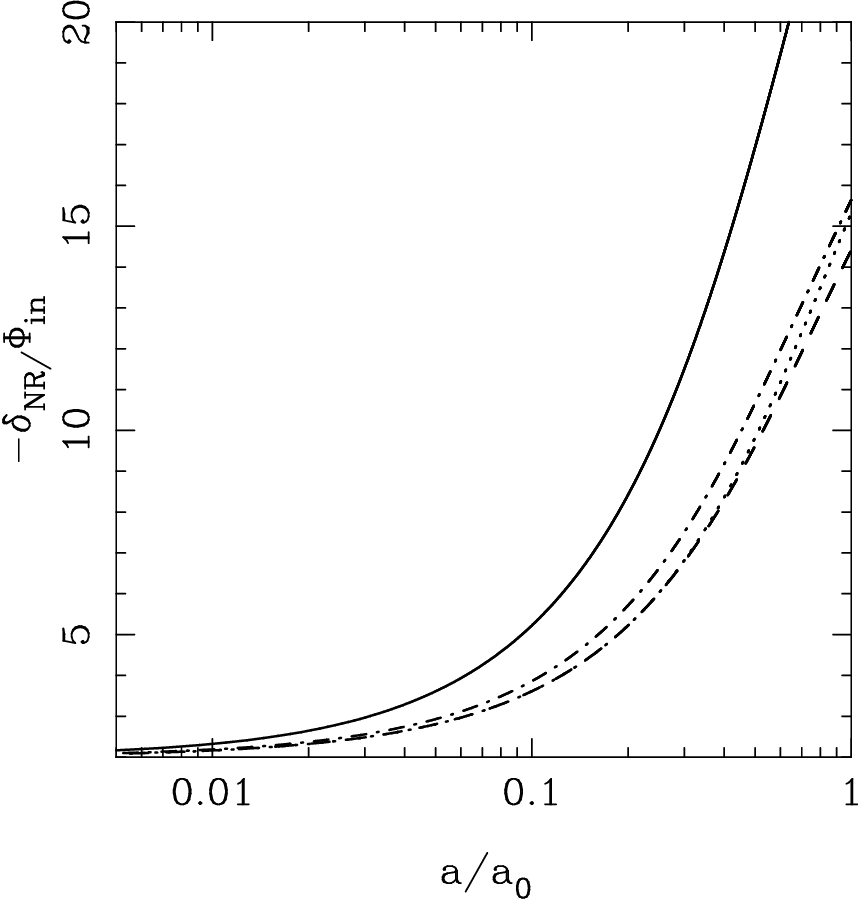}
\\
\includegraphics[width=0.5\textwidth]{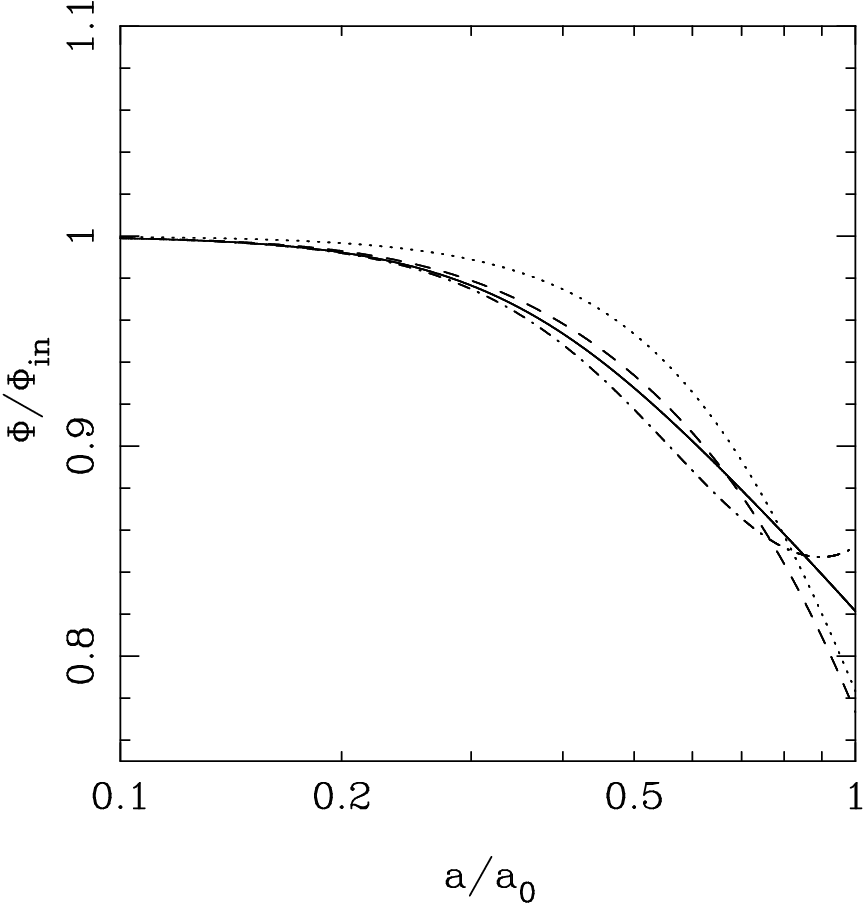}
\includegraphics[width=0.5\textwidth]{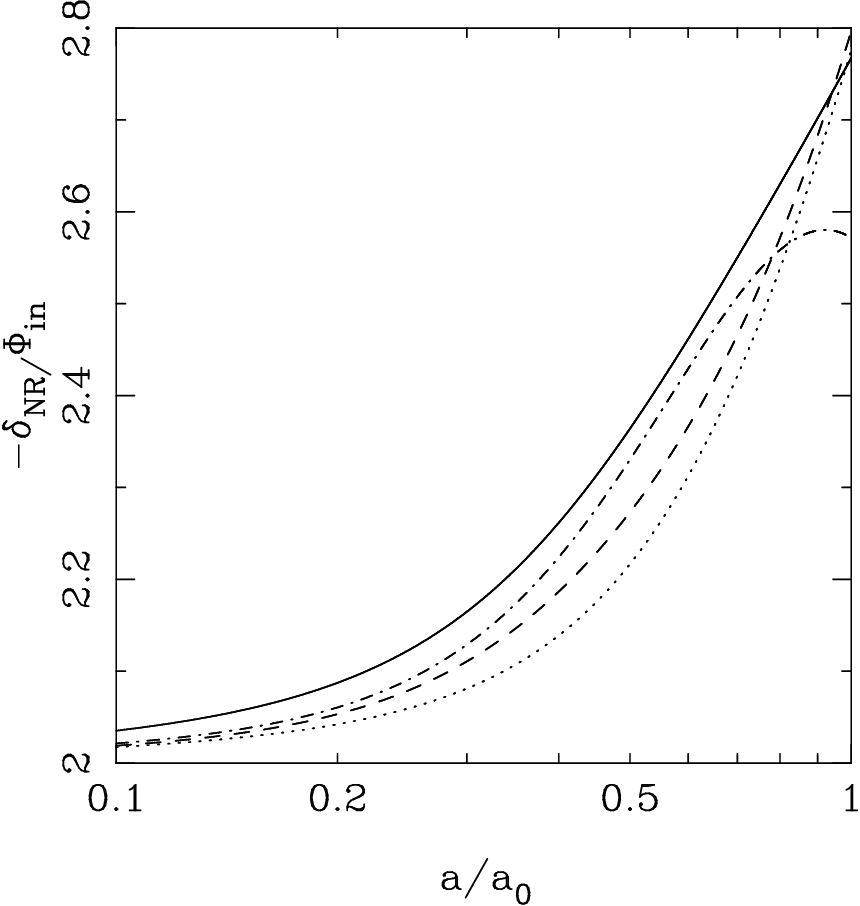}
\end{tabular}
\end{center}
\caption{The upper panel in this figure shows evolution of the gravitational
  potential $\Phi$   and  matter density contrast $\delta_{NR}$ as a
  function of scale factor for the  four different models at length scale
  $\lambda=10^4$ Mpc. The line styles used for different models are  same as
  in Fig. 5. At early times, the
freezing potentials have matter perturbations enhanced as compared to the
thawing potentials. The lower part of the figure shows the evolution of
gravitational potential and matter density contrast at length scale $\lambda =
10^5$.}  
\label{fig::delta}
\end{figure*}

In Fig. \ref{fig::delta1} and Fig. \ref{fig::delta}  we show evolution of
matter perturbations in different models considered here at different length
scales. 
At small scales, say at $\lambda=50$ Mpc, the evolution is indistinguishable
from that in a homogeneous model of dark energy. 
The evolution of perturbations in different in different scalar field models
but the limit of a smooth dark energy works well at small length scales.
At this scale, except for the inverse power law potential model, evolution in
the scalar field models closely follows that in the cosmological constant
model. 
The fluid approximation too works well at these scales.
Explicit dependence of dark energy perturbations is significant at large
scales.

At scales larger than the Hubble radius, the evolution of perturbed quantities
are significantly different from those in homogenous dark energy model. 
In Fig. \ref{fig::delta}, we plot the gravitational potential and matter
density contrast at length scales $\lambda=10^4$ and $\lambda=10^5$ Mpc.
The freezing models of scalar field, at early times have significant
enhancement in matter perturbations at early times. 
There is a significant enhancement of matter perturbations as compared to the
other models. 
This enhancement is more for models with higher $n$.
Among the scalar field models considered here, models FR1 have the highest
rate of growth of perturbations.  
The dashed line denotes FR2, with $n=2$ and $\alpha=0.8$. 
As compared to the cosmological constant, FR2 has a higher rate of growth of
nonrelativistic matter density contrast. 
The dot dashed line corresponds to the potentials TH2 with $n=2$. 
However, the quadratic potential shows a larger growth rate at early times and
then shows a downward trend when the scalar field approaches the minimum of
the potential. 
At early times, TH2 has more enhancement in matter density contrast than the
FR2 potential.   
This is because the dark energy equation of state deviates from that
of a cosmological constant for the polynomial potential (TH2) more than for the
inverse power exponential potential (FR1).
At late times, for TH2 the scalar field reaches the
minimum of the potential and the equation of state oscillates between $-1$ and
$+1$ and the perturbations in dark energy and consequently matter
perturbations begin to oscillate in future.
The late time suppression shown in the figure is due to the beginning of the
oscillatory phase.
In the case of an exponential potential (TH1), matter density contrast is
suppressed compared to all other models.

\section{Conclusions} \label{sec::concl}

In this paper, we analyze the growth of perturbations in scalar field dark
energy scenarios.
The assumption that the distribution of dark energy (with $w \neq -1$) is
homogeneous at all length scales is inconsistent with the equivalence
principle.  
On length scales comparable to or greater than the Hubble radius, the
perturbations in dark energy can become comparable to perturbation in matter
if $w_{de} \neq -1$.   
For scales smaller than the Hubble radius, perturbations in dark
energy can be neglected in comparison with the perturbation in
matter at least in the  linear regime.
Hence any deviations from the cosmological constant model can be explored by
assuming dark energy as a homogeneous component and the scalar field models can
be approximated well by parameterized fluid models.
However, on much larger scales, if the equation of state parameter deviates
from -1, then perturbations in dark energy do influence  matter power
spectrum.

For canonical scalar fields, a clustering dark energy  enhances matter
perturbations as compared to the corresponding homogeneous dark energy
scenario.  
This enhancement more pronounced at scales larger than the Hubble radius.
In particular, we study two broad classes of canonical scalar field models,
namely the thawing and freezing models.
The equation of state of dark energy evolves differently in these two classes
of models.
This affects the growth of perturbations in these different types.
For fast rolling, freezing type models, the dark energy equation of state
deviates away from that of a cosmological constant at early times and freezes 
to $w=-1$ at late times.
In slow roll thawing models, the scalar field remains at a constant $w=-1$
and starts to  deviate away from this value at late times.
The present day observations, which are based on distance measurements, cannot
distinguish between these models.

We studied evolution of matter perturbations in the presence of a clustering
dark energy.
The models we have considered in this paper are within the range allowed by
distance measurement observations.
For  corresponding homogeneous dark energy models and clustering dark energy
models, there are significant changes in the matter density contrast
evolution.
All the canonical scalar field models studied here show an enhancement in
matter perturbations if dark energy is perturbed.
Although we have not considered phantom like models in this paper, it is worth
mentioning that in these models, dark energy perturbations suppress matter
perturbations.

In general, freezing type models have a higher rate of growth of density
contrast at early times.
This is due to the fact the in these models, the equation of state of dark
energy is further away from that of a cosmological constant.
Hence dark energy perturbations play a prominent role at earlier times.
For thawing type models, in general, dark energy perturbations affect the
matter perturbations at late times.
For most of the evolution, the matter perturbations remain close to those in
cosmological constant model and being to deviate as the field begins to thaw.
There are significant deviations in the way density contrast grows, not only
between various  models but also  from the concordant cosmological constant
model.   
Apart from  different way matter perturbations grow in the freezing and thawing
classes of models, models within the same class also have variation in the
behavior of the density contrast.
Observable changes in the angular power spectrum at large scales are limited
by the cosmic variance and therefore CMB data alone will be insufficient to
distinguish between these models. 
Since the scales at which dark energy perturbations are relevant are large,
the primary contribution to CMB anisotropies is through the ISW
effect. 
Therefore it is important to study the ISW cross correlation with large scale
structure indicators.

\section*{Acknowledgment}
The author thanks Department of Science and Technology, India for financial
assistance through project number SR/WOS-A/PS-11/2006.

\end{document}